\documentstyle[aps,preprint]{revtex}
\begin{document}
\title{
       Magnetism of small transition metal clusters and the effects
       of isomerisation}
\author{Peter Borrmann, Bernd Diekmann, Eberhard R. Hilf}
\address{Department of Physics, Carl v.~Ossietzky University Oldenburg,
D-26111 Oldenburg, Germany}
\author{David Tom\'anek}
\address{Department of Physics and Astronomy, and
         Center for Fundamental Materials Research,
         Michigan State University,
         East Lansing, Michigan~48824-1116, USA}
\maketitle
%
%    A B S T R A C T
%

\vspace{-0.8cm}

\begin{abstract}
We investigate the magnetic properties of small transition metal clusters
using a simple statistical model, which requires some input
data from {\em ab initio} spin-density functional calculations. In our
study, we consider a thermodynamically equilibrated ensemble of
clusters with different structures, spin multiplicities, and
ground state energies.
We calculate the physical properties of this system by weighting the
individual configurations according to the
Boltzmann statistics. We find that presence of isomers with very similar
ground state energies, yet very different magnetic properties,
gives rise to a rich magnetic behaviour of the
system which differs significantly from what would be expected for single
configurations.
We apply the present model to determine the magnetic susceptibility of
a cluster ensemble of Langevin paramagnets.
Our results show that some of the anomalies in the
magnetic behaviour of transition metal clusters might be understood in
the framework of our model which is, of course, limited by
the extremely high computational effort needed to obtain the input data.
\end{abstract}
\pacs{}
%
%    I N T R O D U C T I O N
%
\narrowtext
\section{Introduction}
The magnetic properties of small transition metal clusters have been of
growing interest in the past few years. However, there has not been any
study about the dependence on temperature so far.

As shown in a large variety of papers (e.g. Refs.\ \onlinecite{borr1,borr2})
small clusters exhibit, unlike the bulk, smooth structural transformations,
which one might call {\em isomer hopping}, and which occur between pure solid
and fluid phases over a relatively wide temperature range, in a
so-called coexistence phase. In the following we will be concerned with the
effect of those properties on the magnetic behaviour of clusters.
The basic idea is that a cluster of a specific size might have
two or more structures with different magnetic moments and that these
structures or {\em isomer states} occur
with their statistical probability.\footnote{
In the following we will view different configurations, even
if they have the same symmetry and only different lattice constants,
as different isomers.}
As shown below, this simple assumption causes a strong deviation
of the paramagnetic behaviour of the magnetic susceptibility from
the Curie law. As an interesting feature, a dependence on
the strength of an external magnetic field might occur, even at
relatively small fields.
Transition metal clusters seem to be natural candidates for
the occurrence of such effects
because of their different possible electron configurations
with very similar energies, which
in turn yield quite different average magnetic moments.
%
%    Theoretical model
%
\section{Theory}
We consider to have $n$ isomers with  magnetic moments $\mu_{i}$ and
ground state energies E$_{0}(i)$, calculated using an appropriate
spin-density functional Hamiltonian.
In the presence of an external magnetic field the energy changes.
First order quantum mechanical perturbation theory yields
\begin{equation}
{\rm H}_{{\rm m}} = g \mu_{\rm B} M B_{\rm ext} + \frac{e^{2}}{8 m_{\rm e}}
B_{\rm ext}^{2} \sum_{\nu}  \langle x_{\nu}^{2} + y_{\nu}^{2} \rangle \; ,
\end{equation}
where g is the Land\'e factor and the magnetic field is pointing in the
$z$-direction (see Refs.\ \onlinecite{weissbluth,kopitzki}).
The first term describes a paramagnet, which will
be considered exclusively in what follows and can be described with
a purely classical description \cite{ludwig}, as
\begin{equation}
E_{\rm m} = - \vec{\mu} \cdot \vec{{\rm H}}          \;.
\end{equation}
Now we are able to write down the partition function as
\begin{eqnarray}
Z &=& \sum_{i}^{n}  \int_{-1}^{1}  {\rm d}\cos\theta
\exp( -\beta E_{0}(i) +  \beta \mu_{i} H \cos(\theta) ) \nonumber \\
\label{part}
&=& \sum_{i}^{n} \exp( -\beta E_{0}(i) )  \frac{2}{\beta \mu_{i} H}
\sinh( \beta \mu_{i} H )                                   \; .
\end{eqnarray}
Indeed this is an extremely simple partition function, which
can be improved to any complexity by adding all degrees of freedom
of the clusters, e.g. by substituting the summation over the isomers
by integration over the whole configuration space over the
atomic positions or by adding a summation over different spin
multiplicities of the clusters.
 From Eq.~(\ref{part}) the average magnetic moment $\langle \mu \rangle$
and the magnetic susceptibility $\chi$
can be calculated from Eq.~(\ref{part}) easily \cite{ludwig}, using
\begin{eqnarray}
\langle \mu \rangle &=&\frac{1}{\beta}
\frac{\partial}{\partial {\rm H}}
\ln {\rm Z}  \nonumber \\
&=&
\frac{ \sum_{i} \exp(-\beta E_{0}(i)) \cosh(\beta \mu_{i} H )}
 { \sum_{i} \mu_{i}^{-1} \exp(-\beta E_{0}(i)) \sinh(\beta \mu_{i} H )}
-  \frac{1}{\beta {\rm H}} \;,
\end{eqnarray}
\begin{equation}
\chi = \lim_{H \rightarrow 0} \frac{\langle \mu \rangle  N_{\rm m}}{V H} \; ,
\end{equation}
if $N_{\rm m}/V$ is taken as the number
of particles per unit volume.\footnote{
The thermal property related to $\chi$ by the dissipation-fluctuation theorem
is $ \gamma = - \frac{1}{\beta^2} \partial_{\rm HH}\ln Z =
\partial_{\rm H} \langle \mu \rangle$.}
In the case of small magnetic fields $\mu$ H $\ll  k_{\rm B} T $
we get
\begin{equation}  \label{avmu}
\langle \mu \rangle \approx  {\rm Z}_{0}^{-1}(\beta) \;
\sum_{i}^{n}  \exp(\!-\beta E_{0}(i)\!)
\frac{1}{3} \; \beta \; \mu_{i}^{2} \; H
\end{equation}
and
\begin{equation}  \label{avchi}
\chi \approx \frac{N_{m}}{V}  {\rm Z}_{0}^{-1}(\beta) \;
\sum_{i}^{n}  \exp(\!-\beta E_{0}(i)\!)
\frac{1}{3} \; \beta \; \mu_{i}^{2} \; .
\end{equation}
Here, $Z_{0}(\beta)$ is the partition function defined in
Eq.~(\ref{part}) for zero magnetic field. Eq.~(\ref{avchi}) reveals simply
that $\chi$ is a linear superposition of the susceptibilities of the
individual isomers  weighted by their thermal probability.
%
%      R E S U L T S   A N D   D I S C U S S I O N
%
\section{Results and Discussion}
The general behaviour of the paramagnetic susceptibility  according to
Eq.~(\ref{avchi}) is plotted in Fig.~1 for a hypothetical system with only
two major isomer states with magnetic moments $\mu_{1}$ and $\mu_{2}$
and their energy difference
$\Delta {\rm E} = {\rm E}_{0}(2) - {\rm E}_{0}(1)$.
Unlike a purely paramagnetic behaviour according to the Curie law,
the plots reveal local minima and maxima.
A similar behaviour has recently been
experimentally found by Cowen {\em et al.} \cite{cowen}
for Fe$_{28}$ clusters in supercages of NaY Zeolite.
The locations of the extrema
only depend on the ratio between the
magnetic moments $\mu_{1}/\mu_{2}$
and $\Delta$E.
For a given ratio $\mu_{1}/\mu_{2}$, the position of the minimum
depends in an almost linear fashion on $\Delta$E, as can be seen in
Fig.~2. The extraordinary sensitivity of these results on $\Delta$E is
illustrated by the fact that a change of $\Delta$E by only a few meV moves
the minimum by hundreds of degrees Kelvin.

In Fig.~3 the average magnetic moments of the canonical ensemble
are plotted for various magnetic fields. These plots reveal
the fact, that in the presence of a magnetic field the occupation
probability of the isomers changes dramatically due to the additional
magnetic energy.

If one views H to be small in the sense of Eq.~(5), the magnetic
susceptibility can be easily calculated. For varying magnetic fields
this calculation yields a behavior which is very similar to that of a varying
ground state energy difference (see Fig.~4).

At first sight, configurations of different symmetry -- such as
bcc versus fcc structures -- of Fe clusters seem to be
the best candidates to show the effects discussed above.
Even if the energy difference between two isomers might
be very small, the transition from one isomer to the other
may involve a complex concerted motion of atoms and may be associated with a
nonzero activation energy. Thus a thermal relaxation might take quite a
long time, and a hysteresis might be observable.

Lee and Callaway \cite{lee} have found interesting results for
Cr$_{9}$ and V$_{9}$  bcc clusters, which are given in Table~I.
With varying lattice spacing the average magnetic moment changes
by a factor of up to five, whereas the ground state energy changes
are only $\approx 0.04$~Ry/atom.
This is especially interesting since even a simple spatial  expansion of the
transition metal clusters intuitively is more probable than a structural
transformation.

 From our results we infer that it is extremely hard
to obtain phenomenological
results for the magnetic behaviour of clusters from
spin-density
functional
methods, since the results depend
very sensitively on ground state energy differences, which are often very small
and comparable in magnitude to the precision of these {\em ab initio} methods.
In addition, for larger clusters, the number of relevant
isomers increases dramatically.

Nevertheless, we have shown that in the model case of
two cluster isomers, the rich
behaviour of the magnetic susceptibility
as a function of temperature and the external
magnetic field is a sensitive tool to probe
structural properties of clusters. On the other hand,
once the general magnetic response
due to the above discussed effects is understood,
one might use the susceptibility of the
clusters as a very sensitive thermometer.

Moreover, by applying a sufficiently strong magnetic field,
one might be able to suppress one
of the transition metal cluster isomers, provided the clusters
have ample time for structural rearrangement.
An extended study of these effects, including specific examples,
will be published elsewhere.

%.........................................................................

\section*{ Acknowledgements }

We thank Professors Cowen and Dye for making their experimental data available
to us prior to publication. One of us (DT) acknowledges financial support by
the National Science Foundation under Grant Number PHY-92-24745 and the Office
of Naval Research under Grant Number N00014-90-J-1396.

%...........................................................................
%
%      R E F E R E N C E S
%

%
%      T A B L E S
%
\begin{table}[htb]
\caption{Average magnetic moments $\mu$ ($\mu_{\rm B}$/atom)
and total energy per atom E$_{\rm total}$ (Ry) for Cr$_{9}$ and V$_{9}$
clusters, as listed in Ref.\ \protect\onlinecite{lee}.}
\begin{tabular}{llllll}
\multicolumn{3}{c}{Cr$_{9}$}& \multicolumn{3}{c}{V$_{9}$}\\
 a/a.u.  & $\mu$ & E$_{total}$ &a/a.u.& $\mu$ & E$_{total}$ \\
\hline
 3.82    & 0.00  & $-2082.556$ & 4.57 &  0.33 & $-1882.202$ \\
 4.10    & 0.00  & $-2082.598$ & 5.14 &  0.33 & $-1882.219$ \\
 4.36    & 0.67  & $-2082.779$ & 5.54 &  0.33 & $-1882.218$ \\
 4.63    & 0.67  & $-2082.724$ & 5.71 &  0.33 & $-1882.213$ \\
 4.90    & 0.67  & $-2082.666$ & 6.28 &  0.33 & $-1882.175$ \\
 5.30    & 0.67  & $-2082.651$ & 6.85 &  2.78 & $-1882.175$ \\
 5.45    & 3.78  & $-2083.011$ &      &       &             \\
 6.00    & 3.78  & $-2082.934$ &      &       &             \\
\end{tabular}
\end{table}
%
%      F I G U R E   C A P T I O N S
%
\begin{figure}
\caption{Magnetic susceptibility $\chi$ /
$\mu_{\rm B}$ $ \frac{\rm N_{\rm m}}{V}$
for an ensemble consisting of two magnetic {\em configuration} states
($\mu_{1} = $1.0 $\mu_{\rm B}$ , $\mu_{2} =$10.0
$\mu_{\rm B}$). $\Delta$E is the energy difference between the two states.}
\end{figure}
\begin{figure}
\caption{Location of the local minima of the magnetic susceptibility as a
function
of $\Delta$E for various ratios $\mu_{1}/\mu_{2}$ of the magnetic moments. }
\end{figure}
\begin{figure}
\caption{Average magnetic moments at various magnetic fields H,
for $\Delta$E = 0.01~eV, $\mu_{1}$ = 1.0~$\mu_{\rm B}$, and
$\mu_{2}$ = 10.0~$\mu_{\rm B}$.}
\end{figure}
\begin{figure}
\caption{ Magnetic susceptibility $\chi$/ $\mu_{\rm B}$
$ \frac{\rm N_{\rm m}}{V}$
at non-zero magnetic fields H (eV/ $\mu_{\rm B}$),
for $\Delta$E = 0.01~eV, $\mu_{1}$ = 1.0~$\mu_{\rm B}$ and $\mu_{2}$ =
10.0~$\mu_{\rm B}$.}
\end{figure}
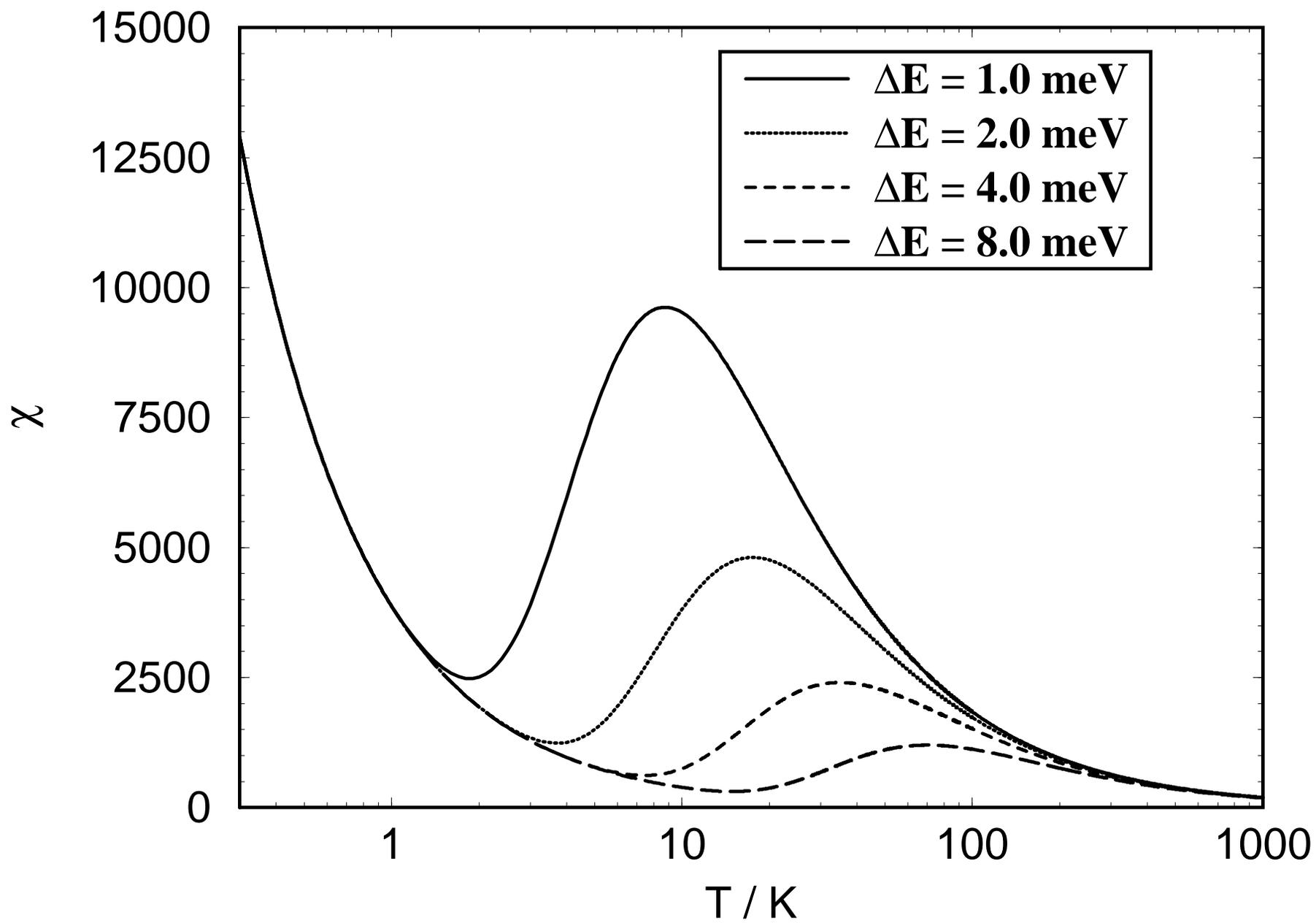
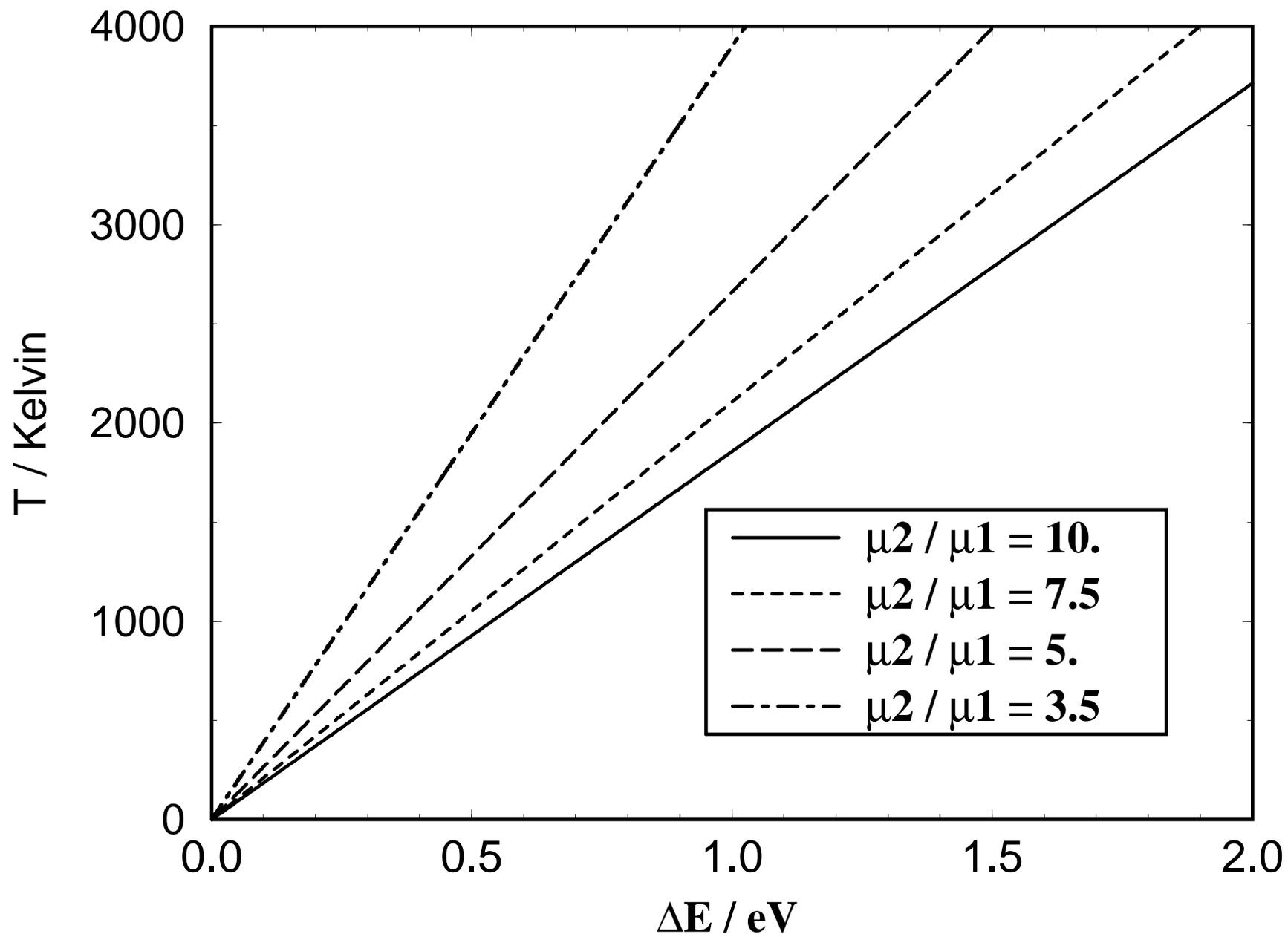
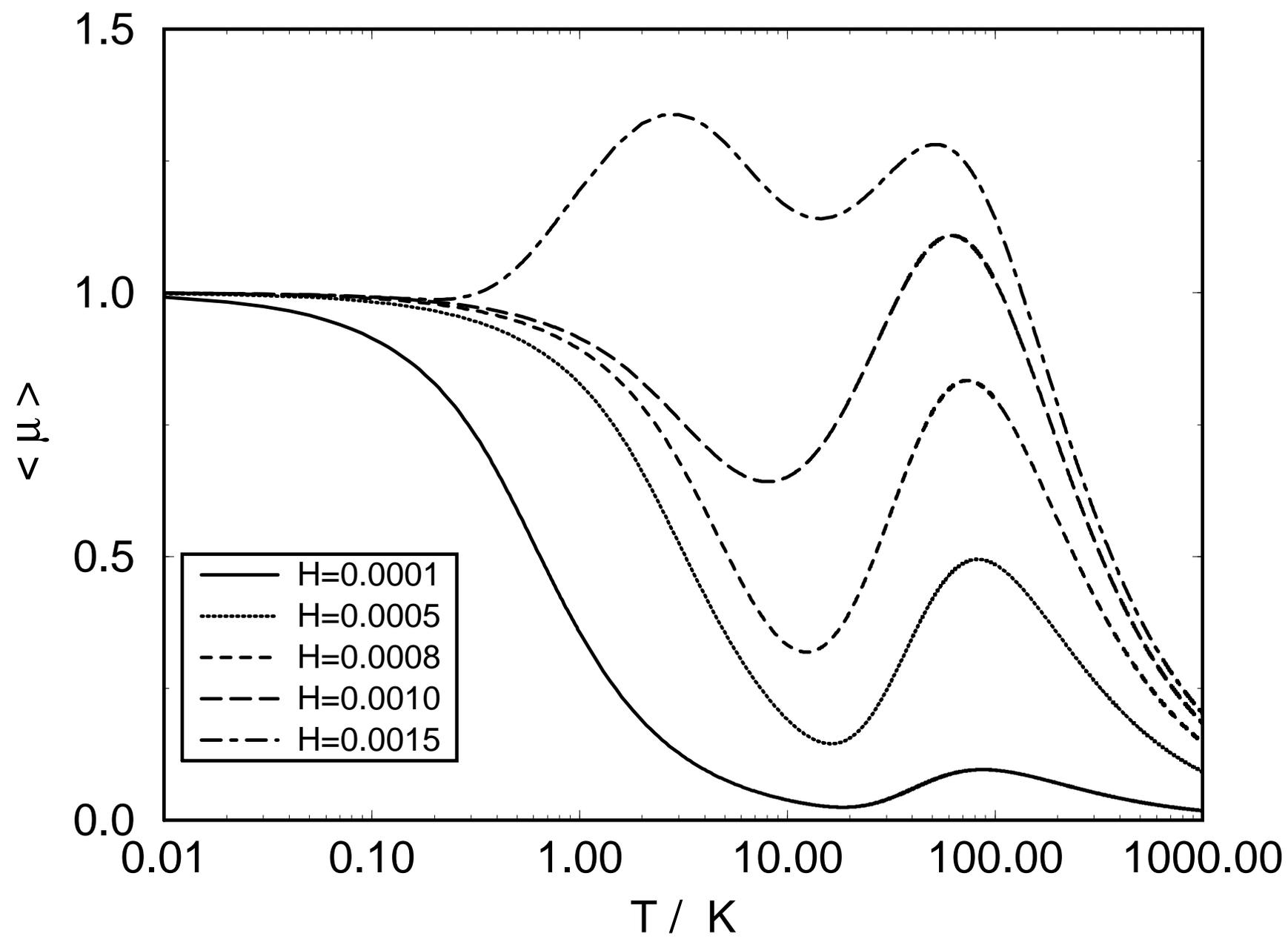
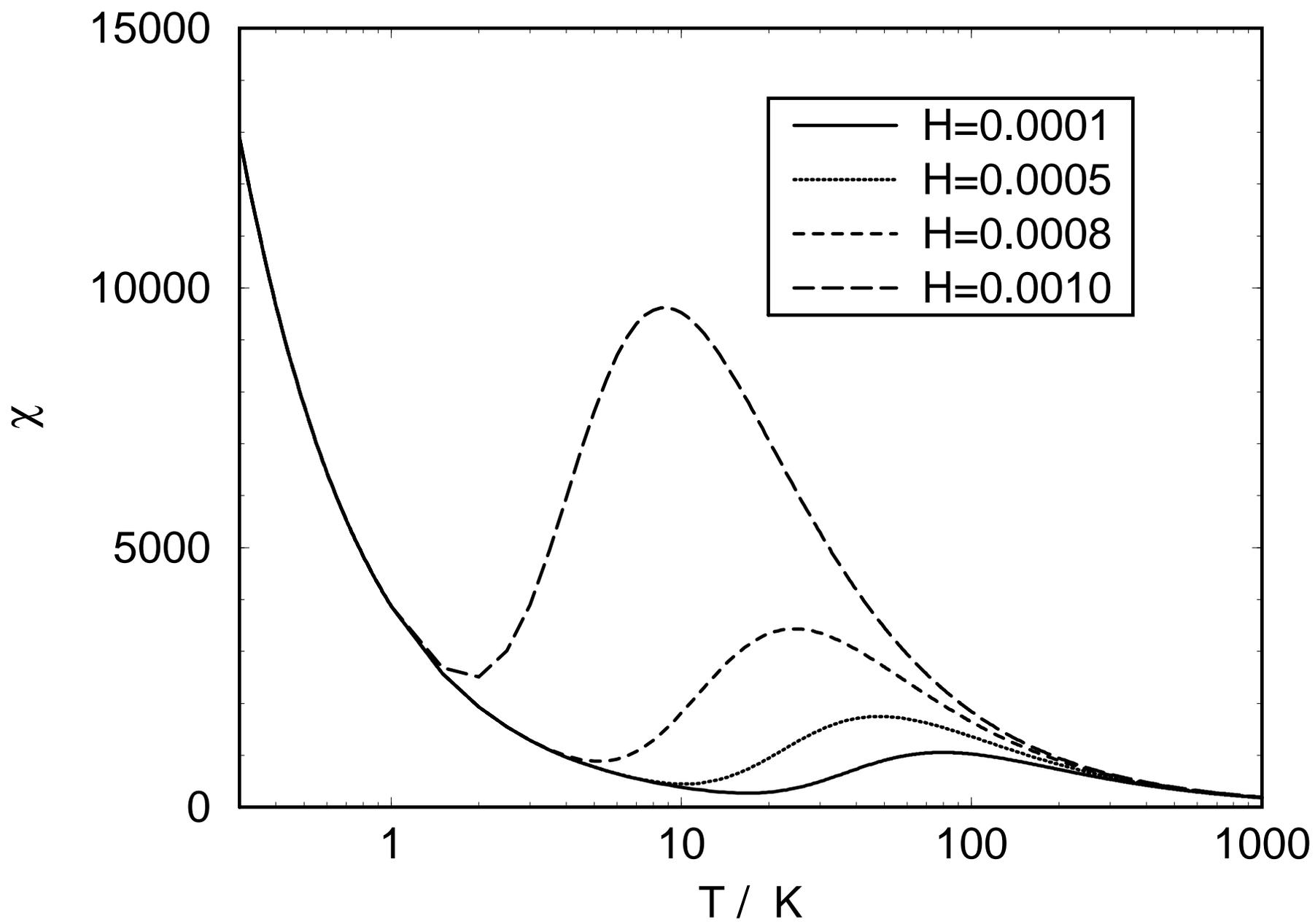
\end{document}